\documentclass[prl,aps,twocolumn,showpacs,floatfix]{revtex4}
\usepackage{graphicx}
\usepackage{bm}
\usepackage{amsmath}
\usepackage{amssymb}
\usepackage{amsfonts}
\usepackage{color}
\usepackage {ulem}
\usepackage{dcolumn}%
\usepackage{bm}
\usepackage{units}
\usepackage{textcomp}

\newcommand\Rey{\mbox{Re}}  
\newcommand\Wi{\mbox{Wi}} 

\newcommand\al{\textit{et~al.}\ }

\newcommand\s{\dot{\gamma}}

\def\mum{\nobreak\mbox{$\;$\textnormal{\textmu m}}}
\def\mlmin{\nobreak\mbox{$\;$\textnormal{\textmu l/min}}}

\begin{document}

\preprint{APS/123-QED}

\title{Geometric scaling of purely-elastic flow instabilities}

\author{J. Zilz$^1$, R. J. Poole$^2$, M. A. Alves$^3$, D. Bartolo$^1$, B. Levach{\'e}$^1$ and A. Lindner$^1$}
\affiliation{$^1$ PMMH UMR7636-ESPCI Paristech-CNRS-Paris 6-Paris 7,
10, rue Vauquelin F-75231 Paris Cedex
05, France.\\
$^2$ School of Eng., University of Liverpool, Brownlow Hill,
Liverpool, L69 3GH, UK.\\
$^3$ CEFT, DEQ, FEUP, Rua Dr. Roberto Frias, 4200-465 Porto, Portugal}

\date{\today}

\begin{abstract}

We present a combined experimental, numerical and
theoretical investigation of the geometric scaling of the onset of a
purely-elastic flow instability in a serpentine channel. Good
qualitative agreement is obtained between experiments, using dilute
solutions of flexible polymers in microfluidic devices, and
two-dimensional numerical simulations using the UCM model. The
results are confirmed by a simple theoretical analysis, based on the
dimensionless criterion proposed by Pakdel-McKinley
\cite{Pakdel1996, McKinley1996} for onset of a purely-elastic
instability.

\end{abstract}

\maketitle


Purely-elastic flow instabilities occur for low Reynolds number
flows of viscoelastic fluids. Such elastically-driven instabilities
- which are entirely absent for the equivalent Newtonian fluid flow
- have been experimentally observed in a range of macroscopic flow
geometries such as viscometric Couette or plate-plate devices
\cite{Larson1990,Shaq1996}. The onset of instability has been
attributed to a balance between normal stresses and streamline
curvature \cite{Shaq1996}. Pakdel and McKinley (Pak-McK)
\cite{Pakdel1996, McKinley1996} proposed an elegantly simple
dimensionless criterion to unify the experimental observations up to
that date. However, beyond the lid-driven cavity results used in
\cite{Pakdel1996}, systematic studies of the instability threshold
as a function of the curvature of a given flow are still lacking. In
this Letter we investigate the geometric scaling of the onset of a
purely-elastic instability in serpentine channels of rectangular
cross-section, combining experiments, numerical simulations and a
simple theoretical analysis. In such wavy channels the geometric
influence on the curvature is dominant and the curvature is
essentially independent of flow rate.   A major complication of
investigating curvature effects in most other complex flows - e.g.
abrupt contractions or lid-driven cavity flows - is that the
dominant flow curvature is not solely determined geometrically but
by the precise dynamics of the flow itself (e.g. varying locally and
with flow rate). These elastic instabilities have proven to be a
good way to obtain efficient mixing in microfluidic devices
\cite{Groisman2001}. Newtonian fluid flow in microfluidic devices
occurs typically in the laminar regime, due to the characteristic
small length scales, and mixing occurs primarily via diffusion, a
very slow process. However, the addition of small amounts of
flexible polymers with high molecular weight is sufficient to impart
strong non-Newtonian behavior, and can even lead to so-called
"elastic turbulence" \cite{Groisman2000}, with increased flow
resistance and enhanced mixing performance \cite{Thomases2009}. It
is thus of great importance to understand the underlying mechanism
that leads to the onset of instability and test the robustness of
the Pak-McK criterion.

\begin{figure}[h!]
  \includegraphics[width=0.8\linewidth]{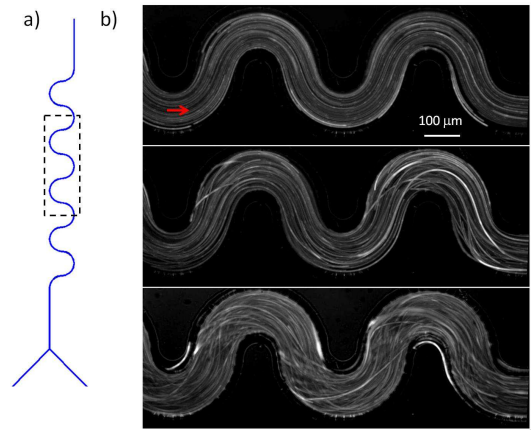}
  \caption{Sketch of the microchannel; typical {\it instantaneous}
  flow patterns in a dilute solution of PEO. From top to bottom: stable
  flow; slightly unstable flow (close to onset of elastic instability); unstable flow.}
  \label{fig_streak_lines}
\end{figure}

We use microfluidic devices,
which are important in applications such as lab-on-a-chip,
fabricated in polydimethylsiloxane (PDMS) from an SU-8 mould, using
standard soft-lithography techniques, which allow straightforward
variation of the flow geometry. Furthermore, due to the small
dimensions of the channel, high shear rates are easily achievable,
thus enhancing the non-Newtonian behaviour observed, while keeping
inertia small. The experiments were performed in serpentine channels
with varying radii of curvature and channel dimensions, using dilute
solutions of a flexible polymer in different solvent viscosities.
The onset of a purely-elastic instability (see Fig. \ref{fig_streak_lines})
was investigated as a function of the
relaxation time of the flexible polymer solution and the geometrical
characteristics of the channel (radius and width). Two-dimensional (2D)
numerical simulations are performed using the upper-convected
Maxwell (UCM) model in identical geometries, assuming that the
curvature in the 2D base-flow is responsible for the instability
(preliminary 3D calculations confirmed the appropriateness of this
approach). The combination of both approaches, together with a
simple theoretical analysis, based on the Pak-McK criterion, allowed
us for the first time to predict the geometric scaling of the onset
of a purely-elastic instability in curved channel flow. In addition
our results independently confirm the strength of the Pak-McK
criterion and open the possibility for extending similar analyses to
a range of other flows.

\begin{figure}
  \includegraphics[width=\linewidth]{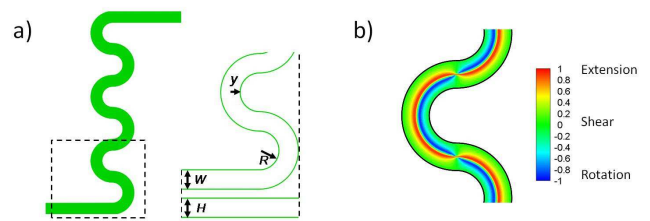}\\
  \caption{a) Geometry and definitions
  ({\it note}: the $y$-direction points to the center along any radial line through each
half loop); b) flow-type parameter. }
  \label{fig_serpentine_flow}
\end{figure}

The geometry of the serpentine channels used in this study are
represented in Fig. \ref{fig_serpentine_flow}(a) (simulations) and
in Fig. \ref{fig_streak_lines} (experiments). The flow takes place
through a given number of half loops in a channel of width $W$ (and
height $H$) and inner radius $R$. In the straight parts of the inlet
and outlet the velocity profile can be estimated using the
analytical solution for fully-developed flow in a rectangular
channel, valid for fluids with a constant shear viscosity (either
Newtonian or viscoelastic). In the curved parts of the channel the
flow becomes weakly asymmetric and the maximum velocity location
occurs slightly closer to the inner wall. In between the two half
loops the flow regains symmetry, before becoming asymmetric towards
the other side in the next half loop. The asymmetry becomes less
pronounced with increasing dimensionless radius $R/W$. The flow is
mainly shear dominated, as can be seen on the contour plot of the
flow-type parameter shown in Fig. \ref{fig_serpentine_flow}b (from
simulations), with limited regions of elongational and rotational
flows close to the centreline where the deformation rates are small.
The flow-type parameter is defined as $\xi = \frac{\arrowvert
\textbf {D} \arrowvert - \arrowvert \boldsymbol \Omega \arrowvert}
{\arrowvert \textbf {D} \arrowvert + \arrowvert \boldsymbol \Omega
\arrowvert} $, where $\arrowvert \textbf {D} \arrowvert $ and
$\arrowvert \boldsymbol \Omega \arrowvert $ represent the magnitudes
of the rate of deformation and vorticity tensors \cite {Lee2007},
and varies from $\xi =-1$ (solid-like rotation), up to $\xi =1$
(extensional flow). A shear-dominated region is identified where
$\xi =0$. We define the average shear rate as $\s=U/W$, with $U$
representing the average velocity in the channel. The Reynolds
number is here defined as $\Rey=\rho U W/\eta$, with $\rho$ and
$\eta$ representing the density and shear viscosity of the fluid,
respectively. The Weissenberg number is defined as $\Wi=\lambda\s$,
where $\lambda$ is the relaxation time of the fluid. While $\Rey$
quantifies the relative importance of inertial over viscous forces,
the degree of elasticity of the flow is quantified using $\Wi$
(ratio of elastic to viscous stresses).

The experiments were performed in serpentine microchannels composed
by eight half loops (Fig. \ref{fig_streak_lines}(a)). We work with
channels of different width, $W=$ 60 and 100$\mum$, but keep the
aspect ratio $a=W/H$ nearly constant and close to 1.4, a typical
aspect ratio for microfluidic devices. The radius of curvature was
varied from $R$=25 to $1950\mum$, in such a way that for each
channel width the dimensionless radius $R/W$ varies approximately
from 0.5 to 20.

Solutions of the flexible polymer PEO (Sigma Aldrich) with a
molecular weight, $M_w = 2 \times10^6$ g/mol, at a concentration  of
125 ppm (w/w) were used in water/glycerol mixtures. The overlap
concentration is $c^* \simeq 860$~ppm \cite{Rodd2007} and we thus
work in the dilute  limit. The solvent viscosity $\eta_s$ varies
from
 $\eta_s=0.93$~mPa.s to 7.4~mPa.s for varying concentrations of glycerol at 23$^\circ$C.
For these solutions the polymer viscosity $\eta_p$ is
approximatively $13\%$ of the solvent viscosity, $\eta_s$, as
measured with a Ubbelohde viscometer. We have also measured the flow
curves of the polymer solutions (not shown) and no shear thinning is
observed. For such small concentrations it is  very difficult to
measure the relaxation time directly \cite{Lindner_PHYSICA_2003},
therefore we estimated $\lambda$ from the Zimm relaxation time.
Since the flow in the serpentine channel is shear dominated, the use
of a relaxation time linked to shear is the most appropriate. From
Rodd \al \cite{Rodd2007} we have calculated the Zimm relaxation time
for PEO of $M_w=2 \times 10^6$~g/mol at 23$^\circ$C in water to be
$\lambda=0.36$~ms. We have fitted the dependence of the Zimm
relaxation time on the solvent viscosity $\eta_s$ as found by Rodd
\al \cite{Rodd2007} (see also \cite{Zell2010}) using a power law
equation, $\lambda=0.4\eta_s^{0.7}$, with $\lambda$ given in ms and
$\eta$ in mPa.s, and use this equation to estimate the relaxation
times of all the solutions used in the experiments.

\begin{figure}
  \includegraphics[width=0.8\linewidth]{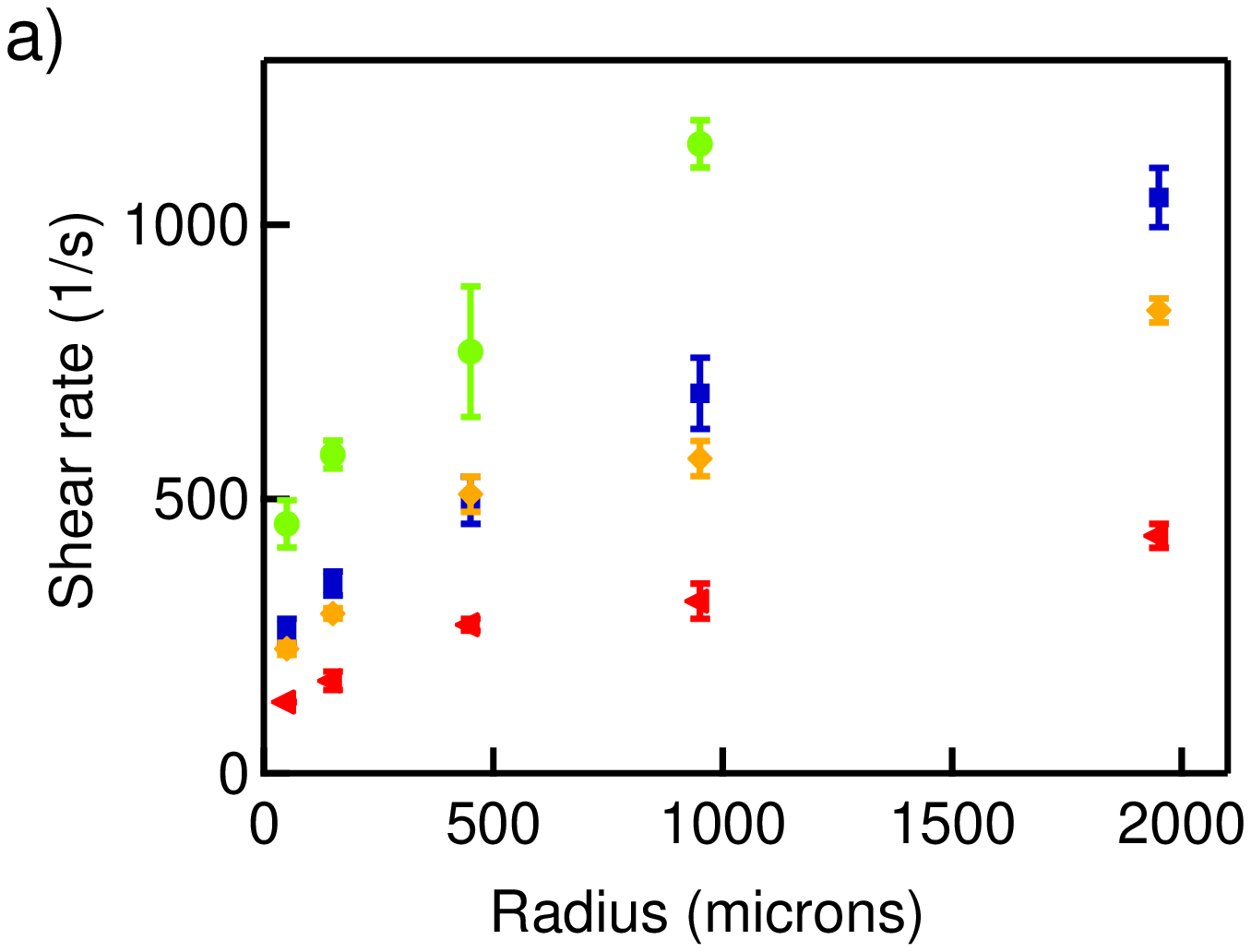}\\
    \includegraphics[width=0.8\linewidth]{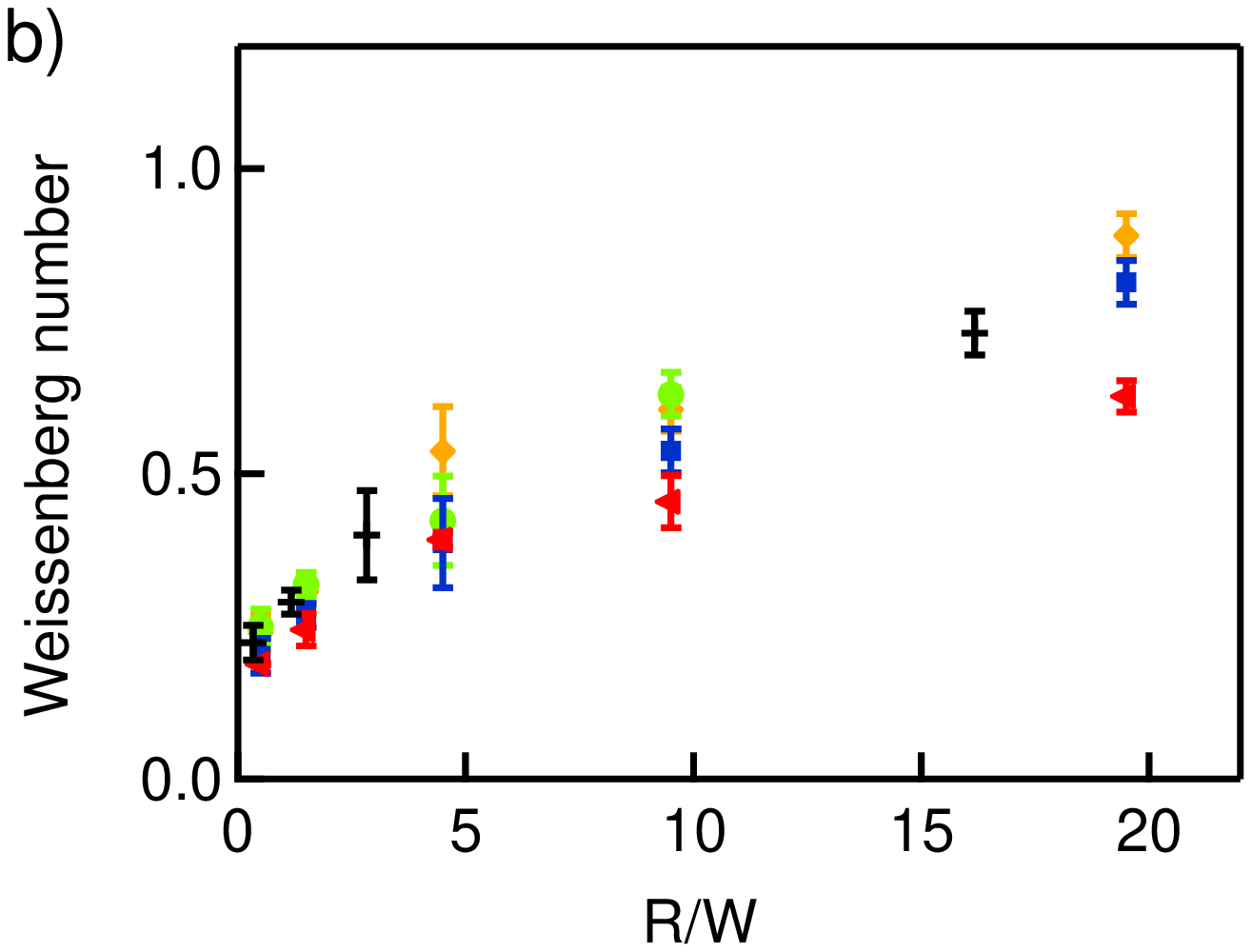}
  \caption{Threshold of instability in a channel of width
$W=100 \mum$ and $H=80 \mum$ for a solution of
  125~ppm of PEO ($M_w=2\times10^6$ g/mol) with different solvent viscosities $\eta_s$ (1.8~mPas $(\textcolor{green}{\bullet})$, 2.8~mPas $(\textcolor{blue}{\blacksquare})$, 4.5~mPas $(\textcolor{yellow}{\blacklozenge})$, 7.4~mPas $(\textcolor{red}{\blacktriangleleft})$) and $W=60\mum$ and
$H=40\mum$ (0$\%$ of glycerol ({\bf --})) (a) Critical shear rate
$\s_c$ as a function of the radius $R$.
  (b) Critical Weissenberg number $\Wi_c$ as a function of $R/W$.}
  \label{fig_Wi_eta}
\end{figure}

The experimental protocol can be summarized as follows. The
solutions are fed to the channel via two inlets, and one of them is
fluorescently dyed. In this way we can visually assess the stability
of the flow. The flow rate ($Q$) was varied from 1  to 50  \mlmin,
and is imposed via a syringe pump (PHD 2000, Harvard apparatus). The
flow is visualized using an inverted microscope (Axio Observer,
Zeiss) coupled to a CCD camera. Starting with the lowest flow rate,
$Q$ is then gradually increased. After each change in $Q$ we wait a
sufficiently long time to achieve steady-state flow conditions (on
the order of 10 minutes). The onset of fluctuations in the flow is
monitored at the last loop of the serpentine channel. This critical
condition defines the onset of the time-dependent elastic
instability, and the critical flow rate $Q_c$ is determined. For
each experimental condition two separate runs are performed using
freshly prepared polymer solutions. We have also visualized the
onset of instability using fluorescent micro-particles (see Fig.
\ref{fig_streak_lines}(b)).

A systematic study was undertaken to determine the critical flow
rate for onset of an elastic instability as a function of the radius
of curvature of the channel, $R$, the width of the channel, $W$, and
the relaxation time of the polymer solutions, $\lambda$. The
critical shear rate, $\s_c=Q_c/(H W^2)$, was determined as a
function of $R$, for a first set of channels of width $W=100\mum$
and height $H=80\mum$, using different solvent viscosities. The
results are depicted on Fig. \ref{fig_Wi_eta}~(a) from which it is
clear that $\s_c$ increases with $R$. Additionally, one can observe
that the critical shear rate is higher for lower solvent
viscosities. The critical Weissenberg number, $\Wi_c=\lambda \s_c$,
is represented in Fig. \ref{fig_Wi_eta}~(b) as a function of $R/W$.
A reasonably good scaling of the data is obtained for all sets of
experiments. Additionally, the data from another set of channels
($W=60\mum$, $H=40\mum$) are also depicted on Fig.
\ref{fig_Wi_eta}~(b) for selected solvent viscosities. Good
agreement between the different channel sizes is obtained, therefore
confirming that $\Wi_c$ is a function of $R/W$. The Reynolds numbers
for the various experiments presented on Fig. \ref{fig_Wi_eta} vary
from 0.2 to 5, therefore, although small, inertial effects are not
negligible. However, we do not observe a modification of the $\Wi_c$
threshold as a function of $\Rey$ (largest for the small solvent
viscosities). For the smallest solvent viscosity and a channel of
radius $R=1950\mum$, the flow does not become unstable in the range
of flow rates accessible in the experiments, and we attribute this
behaviour to a stabilizing effect due to inertia for $\Rey>5$ (as
also observed in \cite{Poole2007}).

For simplicity, the numerical simulations assume 2D planar flow of
an incompressible viscoelastic fluid described by the UCM model. The
equations that need to be solved are those of mass conservation,
$\nabla \cdot \textbf{u} = 0$, and the momentum equation, $-\nabla p
+ \nabla \cdot \boldsymbol \tau = \textbf{0}$, assuming
creeping-flow conditions (i.e. the inertial terms are exactly zero).
The evolution equation for the extra-stress tensor, $\boldsymbol
\tau$, is given by $\boldsymbol \tau + \lambda \boldsymbol
\tau_{(1)}=\eta \dot{\boldsymbol \gamma}$, where $\boldsymbol
\tau_{(1)}$ represents the upper-convected derivative of
$\boldsymbol \tau$. Despite the well-known shortcomings of this
simplified model, such as the unbounded nature of the steady
extensional stresses above a critical strain rate $(\lambda
\dot{\epsilon}=0.5)$, it is the simplest differential constitutive
equation which can capture qualitatively many features of
highly-elastic flows. In addition it forms the backbone of many more
complex models (e.g. the FENE-P, Giesekus and Phan-Thien--Tanner
models), which simplify to the UCM in certain parameter limits, and
thus it is extremely general. The governing equations are solved
using a fully implicit finite-volume numerical method, based on the
logarithm transformation of the conformation tensor, as described in
detail in \cite{Afonso2009} and references therein.

\begin{figure}[h]
  \includegraphics[width=0.8\linewidth]{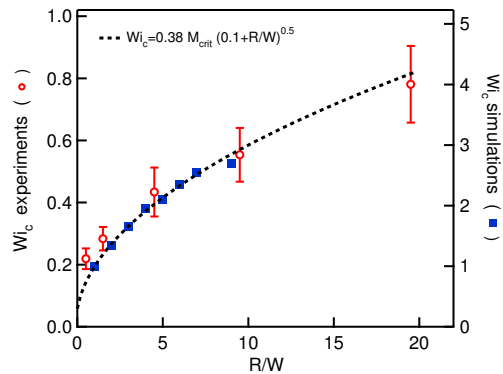}\\
  \caption{Comparison of experimental and numerical results, with the
scaling analysis based on Pak-McK criterion. The experimental data
are the average over the results presented in Fig.
\ref{fig_Wi_eta} (b) for the channel of width $W=100\mum$.}
  \label{fig_Wi_c}
\end{figure}

Fig. \ref{fig_Wi_c} displays the average $\Wi_c$, and the
corresponding standard deviation, of the experimental data, together
with the predictions from the numerical simulations, as a function
of $R/W$. $\Wi_c$ for the simulations reflects the highest $\Wi$ for
which a steady solution could be obtained.  In the experiments the
ratio $\eta_p/\eta_s$ is constant and we can thus expect qualitative
agreement between the experimental data and the results from the
simulations using the UCM model.  This is indeed the case and good
qualitative agreement between experiments and simulations is
obtained.

Pakdel and McKinley \cite{Pakdel1996, McKinley1996} showed that the
curvature of the flow and the tensile stress acting along the
streamlines could be combined in the form of a dimensionless
criterion that must be exceeded for the onset of purely elastic
instability: $\left[\frac{\tau_{11}}{\eta \s}\frac{\lambda u}{{\cal
R}}\right]\ge M^2_{crit}$ with ${\cal R}$,  $u$ and $\s$
representing the local streamline curvature, velocity and shear
rate, respectively. $\tau_{11}$ represents the local streamwise
normal stress and $\eta\s$ the local shear stress. For the UCM model
the normal stress can be written as $\tau_{11}=2\eta\lambda\s^2$.
Assuming fully-developed flow conditions everywhere (below $\Wi_c$),
we can estimate the velocity profile assuming a Poiseuille flow,
$u=\frac{3}{2}U[1-(\frac{y}{W/2})^2]$, with the $y$-axis centred on
the channel and directed along any radial line through each half loop towards the centre.
The shear rate is given as $\s=12U\arrowvert y \arrowvert /W^2$,
therefore one obtains the following criterion for a purely-elastic
instability, $36 (\frac{\lambda U}{W})^2  F(y) \ge M^2_{crit}$,
where $F(y)=\frac {\arrowvert \frac{y}{W} \arrowvert
[1-4(\frac{y}{W})^2]}{\frac{R}{W}+\frac {1}{2}-\frac{y}{W}}$. For
each value of $R/W$ we can determine the location where the onset of
the instability takes place, which occurs at $y_c/W$ where $F(y_c)$
is maximum. Therefore, for each value of $R/W$ we can also determine
the corresponding $\Wi_c$ for onset of instability. In general this
has to be done numerically, but for the limits of very low and very
high values of $R/W$ this can be done analytically, leading to the
following asymptotes and critical locations:

(i) $\Wi_c \rightarrow \frac {M_{crit}} {\sqrt{72}}$; $ y_c/W \rightarrow \frac {1}{2}$ for $\frac{R}{W}
\rightarrow 0$

(ii) $\Wi_c \rightarrow 0.38 M_{crit} \sqrt{\frac{R}{W}} $; $ y_c/W \rightarrow 0.289$ for $\frac{R}{W}
\rightarrow \infty$

These asymptotes can be combined in the general form, $\Wi_c=0.38
M_{crit} \sqrt{ 0.1+\frac{R}{W} }$, which agrees well in the whole
range of $R/W$ with the exact results that are obtained by solving
numerically the general equation derived from the Pak-McK criterion.
When fitting the data with this expression, one obtains
$M_{crit}=2.46$ and $0.48$ for the simulations and the experiments,
respectively. The comparison of the results from experiments,
simulations and the scaling argument thus show that $\Wi_c$ at the onset
of elastic instability in a serpentine channel scales as the square
root of $R/W$ with a small offset when $R/W$ tends towards zero.

\begin{figure}
  \includegraphics[width=0.7\linewidth]{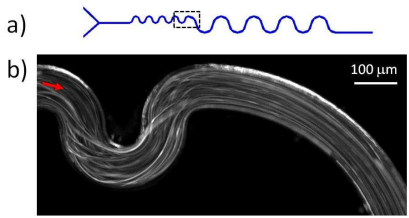}\\
  \caption{a) Sketch of the modified channel; b) streak photography for a
  dilute solution of PEO:
  the flow is unstable in the part of the channel with small radius ($R=50\mum$), but stable in the part with the large radius
  ($R=450\mum$).}
  \label{fig_supercrit}
\end{figure}

We have also tested the nature of the instability in the
experiments. First we have shown that increasing or decreasing $Q$
does not lead to different values of $\Wi_c$ and no hysteresis is
observed. Second, we have perturbed the flow at the inlet using a
modified serpentine channel composed of a series of loops of small
radius $50\mum<R_1<250\mum$, followed by a serpentine channel of
larger radius, $R_2=450\mum$ (see Fig. \ref{fig_supercrit}). No
modification of the threshold in the second serpentine channel is
observed when varying the radius of the first channel. The streak
image shown in Fig. \ref{fig_supercrit} was taken at a flow rate
where the flow in the first channel is unstable whereas the flow in
the second channel is stable. One observes a fast decay of the
perturbation once the fluid enters the second channel. We thus
conclude that the instability in the serpentine channel is
supercritical in contrast to observations made in Couette or
plate-plate devices \cite{Morozov2007}.

Using a combined experimental and numerical approach we investigated
the geometrical scaling of a purely-elastic instability in a
serpentine channel. In this way we have shown that the critical
Weissenberg number at the onset of the instability scales as the square root
of the inner radius divided by the width of the channel, $R/W$, with
a small offset when $R/W \rightarrow 0$, in agreement with a simple
theoretical analysis based on the dimensionless criterion of
Pakdel-McKinley  \cite{Pakdel1996, McKinley1996}.

{\it Acknowledgements} -- We thank A. Morozov and B. Andreotti for
useful discussions, A. Morozov for a critical reading of the
manuscript and P.C. Sousa for rheological measurements.

\end{document}